\documentclass[aps,prl,twocolumn,superscriptaddress,showpacs,10pt]{revtex4-1} 
\usepackage{graphicx} 
\usepackage{amsmath} 
\usepackage{amsfonts} 
\usepackage{color} 
 
\begin{document}

\title{Spin noise spectroscopy of donor bound electrons in ZnO} 

\author{H. Horn}
\affiliation{Institut f{\"u}r Festk{\"o}rperphysik, Leibniz Universit{\"a}t Hannover, Appelstr. 2, 30167 Hannover, Germany}
\author{A. Balocchi}
\author{X. Marie}
\affiliation{INSA-CNRS-UPS, LPCNO, Universit\'e de Toulouse, 135 Av. de Rangueil, 31077 Toulouse, France}
\author{A. Bakin}
\author{A. Waag}
\affiliation{Institute for Semiconductor Technology, Technische Universit{\"a}t Braunschweig, Hans-Sommer-Stra\ss e 66, 38106 Braunschweig, Germany}
\author{M. Oestreich}
\author{J. H{\"u}bner}
\email{jhuebner@nano.uni-hannover.de}
\affiliation{Institut f{\"u}r Festk{\"o}rperphysik, Leibniz Universit{\"a}t Hannover, Appelstr. 2, 30167 Hannover, Germany} 

\date{\today} 

\begin{abstract} 

We investigate the intrinsic spin dynamics of electrons bound to Al impurities in bulk ZnO by optical spin noise spectroscopy. Spin noise spectroscopy enables us to investigate the longitudinal and transverse spin relaxation time with respect to nuclear and external magnetic fields in a single spectrum. On one hand, the spin dynamic is dominated by the intrinsic hyperfine interaction with the nuclear spins of the naturally occurring $^{67}$Zn isotope. We measure a typical spin dephasing time of 23~ns in agreement with the expected theoretical values. On the other hand, we measure a third, very high spin dephasing rate which is attributed to a high defect density of the investigated ZnO material. Measurements of the spin dynamics under the influence of transverse as well as longitudinal external magnetic fields unambiguously reveal the intriguing connections of the electron spin with its nuclear and structural environment. 
\end{abstract} 


\maketitle 

Among the oxide based II-VI semiconductors, ZnO has been devoted a continuously high attention in what concerns its optoelectronic properties \cite{Ozgur.JAP.2005} and possible application in semiconductor spintronics \cite{Zutic.RMP.2004, Dietl.Science.2000}. Especially the long electron spin coherence times at room temperatures which results from the rather weak  influence of the spin orbit splitting onto the conduction band states \cite{Ghosh.APL.2005, Harmon.PRB.2009} makes ZnO and ZnO nanostructures \cite{Janssen.NanoLett.2008} a promising material in semiconductor spintronics and spin based quantum-optronics \cite{Liu.PRL.2007}. The semiconductor material ZnO is easily available with a high abundance and nowadays comes with a vast selection of growth and structuring methods \cite{Janotti.RPP.2009}. Furthermore, ZnO bears a plethora of interesting spin physics due to its intriguing exciton dynamics \cite{Lagarde.PRB.2008} and scalability of its nuclear spin properties \cite{Whitaker.JPCC.2010}.

In this work we investigate the intrinsic spin dynamics of donor bound electrons in the wide band gap material ZnO by optical spin noise spectroscopy via below band gap Faraday rotation. Spin noise spectroscopy measures the omnipresent fluctuations of the spin degree of freedom \cite{Bloch.PR.1946, Aleksandrov.SPJETP.1981, Katsoprinakis.PRA.2007,Crooker.NATURE.2004,Horn.PRA.2011} and has developed into a powerful tool for the investigation of the intrinsic spin dynamics in semiconductors \cite{Muller.PE.2010} since it generally avoids optical excitation \cite{Muller.PRL.2008, Romer.PRB.2010, Dahbashi.APL.2012}. In the measurements presented here, SNS reveals the transverse and longitudinal spin relaxation of donor bound electrons due to hyperfine interaction whereat only a low fraction of the host nuclei carry a nuclear spin. Both, transverse and longitudinal times are acquired in a single spin noise spectrum by SNS which allows the straightforward extraction of the influence of nuclear and external magnetic fields. Furthermore, we observe an additional, very short spin dephasing time which we attribute to the increased interaction with defects located inside the effective donor volume.

\begin{figure}[t]
        \includegraphics[width=\columnwidth]{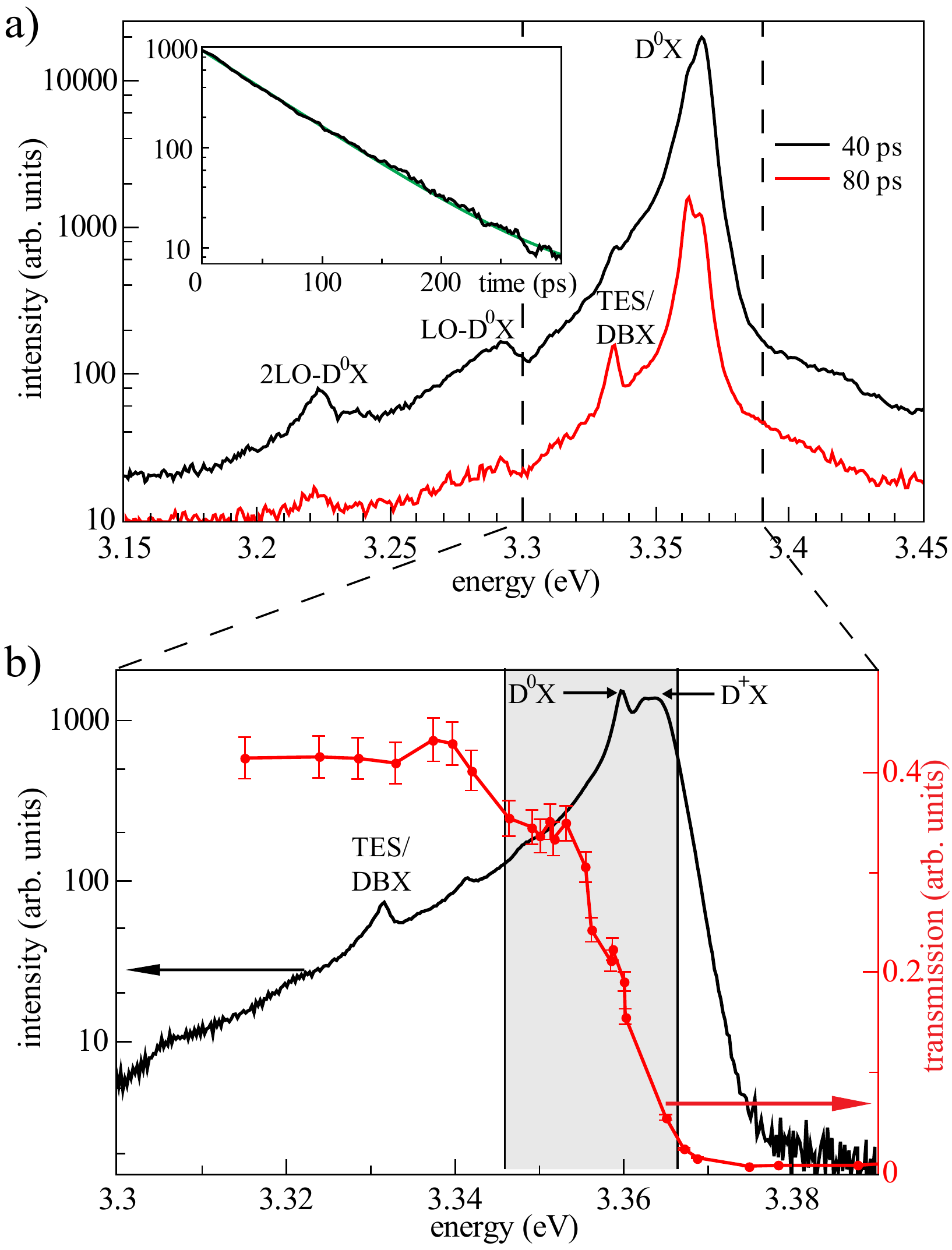} 
        \caption{(Color online) a) Photoluminescence spectra recorded at 40 ps and 80 ps, respectively, after excitation with a picosecond laser pulse with a photon energy of 3.54 eV and a sample temperature of T=4\,K. The inset shows the PL decay transient of the donor bound D$^{0}$X exciton. b) CW-PL (black line) recorded with a higher resolution. The labels D$^0$X and D$^+$X mark the spectral positions of the neutral and ionized donor-bound exciton transition, respectively. The red dots are transmission measurements performed with a spectrally narrow laser.} 
        \label{fig:PLspectra} 
\end{figure} 
The sample is a thin film of predominantly bulk, (0001)-grown ZnO with a nominal thickness of 450~nm deposited on a sapphire substrate by MBE growth. Investigations by scanning electron microscopy reveal a nanoporous structure with a granularity varying between 10 and 20~nm at the surface. Henceforth, the investigated material most likely exhibits a high defect density induced by surface states which is confirmed by a short photoluminescence lifetime at low temperatures which drops even further at elevated temperatures. Figure~\ref{fig:PLspectra}a depicts the time resolved photoluminescence spectrum of the sample recorded with a synchroscan streak camera system under above band gap excitation with a frequency doubled, picosecond Ti:sapphire laser oscillator \cite{Hubner.BOOK.2008}. The donor-bound exciton transition D$^{0}$X and its LO phonon replicas are clearly visible. The inset depicts the time transient of the D$^{0}$X transition decaying mono-exponentially with a lifetime of $\tau_{l}=60$~ps measured at $T=4$~K. The donor-bound exciton transition is the dominant optical transition. The spin dynamics of the donor bound electron is explored in the spin noise measurements presented later. For a better characterization of the donor-bound exciton transition, we performed continuous-wave (cw)-photoluminescence and transmission measurements which are shown in Fig.~\ref{fig:PLspectra}b. From the photoluminescence spectra we attribute the neutral D$^{0}$X (3.360 eV) and ionized D$^{+}$X (3.364) transition to Al impurities besides an unidentified background PL signal around the D$^{+}$X transition \cite{Klingshirn.BOOK.2010}. The shaded area in Fig.~\ref{fig:PLspectra}b indicates the spectral region where SNS measurements are performed in dependence of the probe photon energy detuning relative to the D$^{0}$X transition. These measurements are described further below.

\begin{figure}[t] 
        \includegraphics[width=\columnwidth]{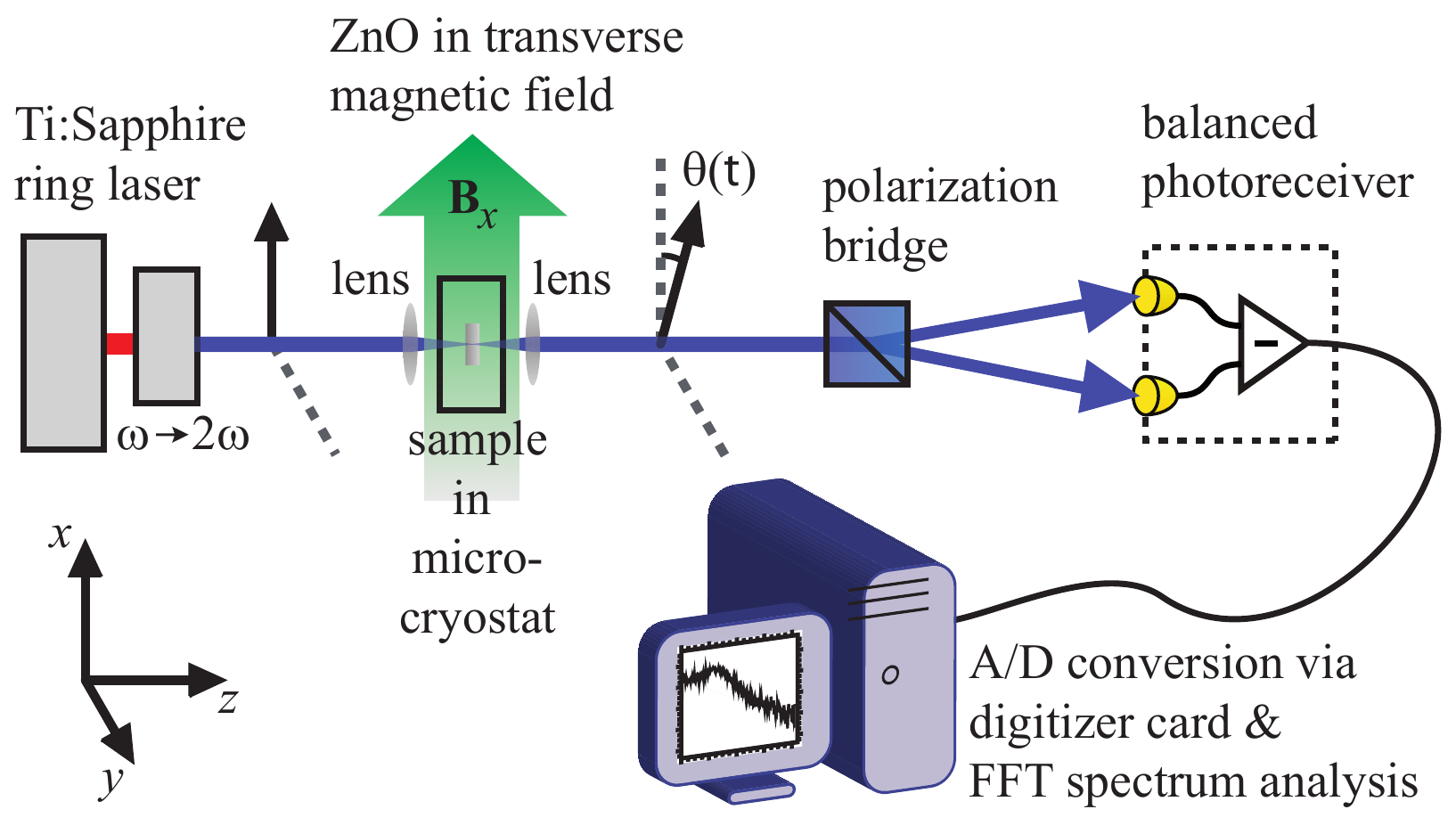} 
        \caption{(Color online) Experimental spin noise setup. The light source is a frequency doubled cw-Ti:Sapphire laser. The linear polarized probe light is focussed through the sample which is mounted in a Helium cold-finger cryostat. The rotation of the linear polarized light is measured by a polarization bridge and a balanced receiver. The electrically amplified difference signal is digitized in the time-domain and spectrally analyzed by a computer. External magnetic field can be applied in transverse and longitudinal (not shown) direction.} 
        \label{fig:setup} 
\end{figure} 
Figure~\ref{fig:setup} depicts the experimental setup for the measurements of the spin dynamics in transmission by spin noise spectroscopy. The light source is a frequency doubled, cw-Ti:sapphire ring laser with a spectral width of $0.3$~neV. The linear polarized light is focused down to a spot size with a typical diameter of $3~\mu\rm{m}$ and an intensity of $132~\mu\rm{W}/\mu\rm{m}^{2}$ and transmitted through the sample. The sample is mounted in a Helium cold finger cryostat and cooled  to a temperature of 4~K. The transmitted probe beam acquires the stochastic spin dynamics of the donor electron spin ensemble in the ZnO sample via below band gap Faraday rotation. The time-dependent, fluctuating Faraday rotation angle is analyzed by a polarization bridge consisting of a polarizing beam splitter and a balanced photo receiver. Finally, the time-domain Faraday rotation data is analyzed via fast Fourier transformation (FFT) spectral analysis in the radio frequency regime on a standard personal computer. 
 
\begin{figure}[t]
    \includegraphics[width=\columnwidth]{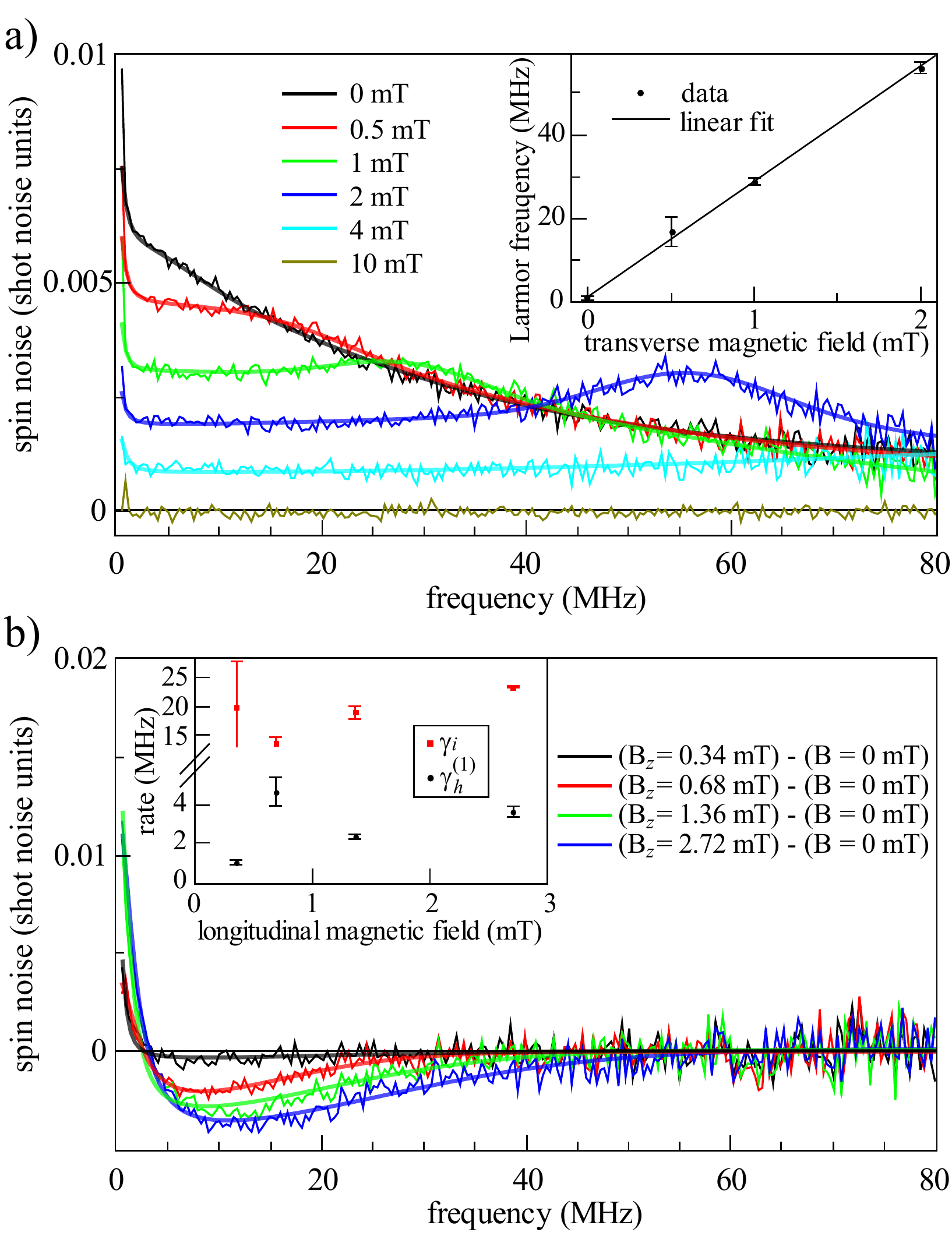} 
    \caption{(Color online) a) Spin noise spectra recorded with different \emph{transverse} magnetic fields. The background spectrum has been acquired at $B_{\perp}=10$~mT and the probe laser photon energy is 3.355~eV. The data is fitted by a model according to the description in the text. The inset depicts the change in Larmor frequency for the inhomogeneous spin noise contribution. b) Spin noise difference spectra for different \emph{longitudinal} magnetic fields. The spectra are normalized to an average spin noise power density in the range of 50 to 80~MHz, in order to compensate for experimental drifts. The data is fitted according the model described in the text. The inset compares the homogeneous longitudinal ($\gamma_{h}^{(1)}$) and inhomogeneous transverse ($\gamma_{i}$) spin relaxation rates in dependence of $B_{\parallel}$. The probe laser photon energy is 3.359~eV for these measurements.} 
    \label{fig:NoisePowSpectrum} 
\end{figure}

The spin noise signal is low compared to the strong background optical shot noise and hence a background noise spectrum is recorded at a transverse magnetic field of $\mathbf{B}_{\perp}=10~\rm{mT}$ which is then subtracted from the lower field spectrum. At $\mathbf{B}_{\perp}=10~\rm{mT}$, the spin noise signal is completely shifted out of the measurable spectrum and hence only optical shot and electrical noise is recorded. Furthermore, the obtained noise spectrum is divided by the optical shot noise spectrum \footnote{The subtraction of the background noise power spectrum with the noise spectrum of the unilluminated detector yields the optical shot noise spectrum. Hereby the resulting spin noise power spectrum is expressed in units of optical shot noise power.} in order to eliminate the non-uniform amplification of the balanced detector. The result is a normalized noise spectrum with only positive values (see Fig.~\ref{fig:NoisePowSpectrum}a). A typical mono-exponential, i.e., homogeneous decay of the spin orientation with a decay constant $\tau_{s}$ results in a Lorentzian shaped spin noise contribution in the noise spectra with a full width given by $\gamma_{h}=1/(\pi \tau_{s})$. Inhomogeneities in the sample broaden the spin noise signal due to, e.g., g-factor variations \cite{Berski.ArXiv.2012} or hyperfine interaction \cite{Merkulov.PRB.2002} and lead to a frequency spectrum following a normal distribution with a standard deviation $\sigma_{s}$. The inhomogeneous spin dephasing rate $\gamma_{i}$ is finally given by $2\sigma_{s} = \gamma_{i}/\pi$ \cite{Pines.PR.1955}. 

The spin noise spectra with applied \emph{transverse} magnetic field $\mathbf{B}_{\perp}$ are shown in Fig.~\ref{fig:NoisePowSpectrum}a and reveal three distinct contributions. First, a very low homogeneous spin dephasing rate $\gamma_{h}^{(1)}$ appearing as a Lorentzian like peak centered at zero frequency. We attribute this contribution to the longitudinal spin relaxation rate of localized electrons with respect to the stochastic nuclear field orientation $\mathbf{B}_{N}$ at the respective donor site. Second, an inhomogeneous spin noise contribution $\gamma_{i}$ originating  from the same localized electrons but which follows the applied transverse magnetic field with the respective Larmor frequency $\nu_{\rm{L}}=g^{*} \mu_{\rm{B}} B / h$\,. The magnitude of $\gamma_{i}$ is given by the dispersion of $\mathbf{B}_{N}$ within the ensemble of localized electrons. A third spin noise contribution with a very high homogeneous spin dephasing rate $\gamma_{h}^{(2)}$ and centered as well at the Larmor frequency $\nu_{\rm{L}}$ is attributed to donor-bound electrons interacting with one or more defects within the effective donor-volume. This is discussed in more detail later. The dependence of the Larmor frequency $\nu_{\rm{L}}$ with magnetic field is shown in the inset of Fig.~\ref{fig:NoisePowSpectrum} and yields an effective Land{\'e} g-factor of $g^{*}=1.97(9)$ which fits very well to the Land{\'e} g-factor of the Al-donor bound electron as measured by electron paramagnetic resonance spectroscopy \cite{Orlinskii.PRB.2008}. This assignment is consistent with the spectral identification of the D$^{0}$X transition as elucidated in the previous section. The width of the inhomogeneously broadened spin noise spectrum, i.e., $\gamma_i$ is determined by the fluctuating nuclear field sampled by the localized electron wave function \cite{Merkulov.PRB.2002}. In ZnO only $\frac{1}{2} \times 4.1\%$ of all lattice ions interact via hyperfine interaction with the donor electrons due the natural abundance of the $^{67}$Zn isotope with a nuclear spin of $I_{N}=5/2$. The electron wave function of an electron bound to an Al-donor has an effective Bohr radius of $r_{B}=1.93$\,nm and thus experiences an interaction with $\sim 5\times 10^{3}$ nuclei. Taking the natural abundance of $^{67}$Zn into account together with a Fermi-contact hyperfine interaction strength of $3.7\,\mu\rm{eV}$ \cite{Whitaker.JPCC.2010} one obtains a standard deviation of the local magnetic field of $\Delta B_{N}=0.22$~mT which corresponds to a theoretical limit for the inhomogeneous spin dephasing time of $(\pi\gamma_{i})^{-1}=26.3$~ns. The extracted inhomogeneous spin dephasing times from the data presented in Fig.~\ref{fig:NoisePowSpectrum}a yield $T_{2}^{*}=23(\pm 2.5)$\,ns matching very well to the theoretical expected times limited by the hyperfine interaction. 

The longitudinal spin relaxation rate $\gamma_{h}^{(1)}$ is defined with respect to the effective magnetic field axis, i.e., $\mathbf{B}_{N} + \mathbf{B}_{\perp}$\,. The corresponding spin noise contribution appears as a narrow Lorentzian like centered at zero frequency. The spin noise power density at very low frequencies is usually superimposed by $1/f$ electrical noise which makes a clear assignment of $\gamma_{h}^{(1)}$ difficult. However, a longitudinal spin relaxation time  $(\pi\gamma_{h}^{(1)})^{-1}\gtrsim 200$~ns can be extracted even without taking into account the frequency range from DC to 500~kHz in the data evaluation. Furthermore, the observed spin noise power associated with $\gamma_{h}^{(1)}$ decreases with increasing transverse magnetic field since the average projection of $\mathbf{B}_{N} + \mathbf{B}_{\perp}$ onto the direction of observation decreases \cite{Glazov.ArXiv.2012}.

\begin{figure}[t]
        \includegraphics[width=\columnwidth]{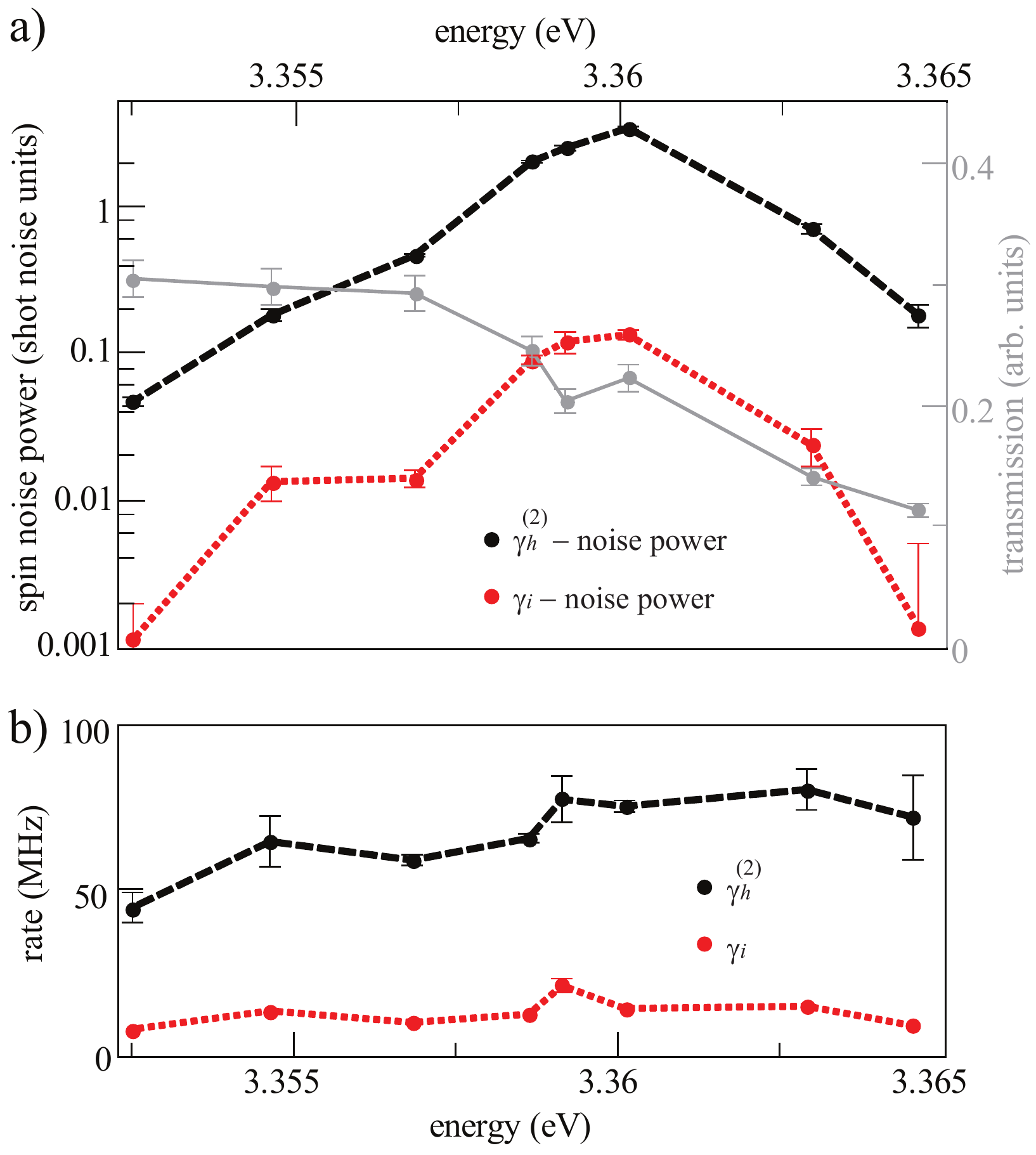} 
        \caption{(Color online) Dependency of a) the total spin noise power and b) spin dephasing rate on the detuning relative to the D$^{0}$X transition. The spectral position of the maximum spin noise power at 3.360~eV indicates that the major spin noise contribution originate from Al-donor bound electrons.}
        \label{fig:LaserTuningDependence} 
\end{figure} 
The situation changes for external magnetic fields applied in \emph{longitudinal}, i.e., $z$-direction. Here an increasing magnetic field quenches the influence of the stochastic nature of the hyperfine interaction \cite{Dahbashi.APL.2012}. Figure~\ref{fig:NoisePowSpectrum}b shows spin noise spectra obtained by measuring the difference between spin noise recorded with an applied longitudinal $\mathbf{B}_{\parallel}$ and zero magnetic field. Clearly visible is that spin noise power from the inhomogeneous transverse part is redistributed to the homogeneous longitudinal spin noise contribution with increasing $\mathbf{B}_{\parallel}$. The total spin noise power is constant in thermal equilibrium, i.e., the external magnetic field is small compared to $k_{\rm{B}}T/\mu_{\rm{B}}$ and polarization effects are negligible. The data presented in Fig.~\ref{fig:NoisePowSpectrum}b is fitted by a single Lorentzian and Gaussian function centered both at zero frequency. We could not resolve the Maxwell-type distribution of the nuclear field orientation \cite{Glazov.ArXiv.2012}. The inset of Fig.~\ref{fig:NoisePowSpectrum}b shows the spin dephasing rates $\gamma_{i}$ and $\gamma_{h}^{(1)}$. The longitudinal spin dephasing rate is slightly higher compared to the measurements with $\mathbf{B}_{\perp}$ due to a smaller detuning from the optical resonance of the D$^0$X in the corresponding measurements. The homogenous spin contribution with the very high spin dephasing rate $\gamma_{h}^{(2)}$ cancels out in the difference measurements with longitudinal magnetic field due to the strongly dispersed spin noise power density.

In order to unveil the origin of the third, very fast homogeneous spin noise contribution $\gamma_{h}^{(2)}$ we perform spin noise measurements with a varying detuning with respect to the D$^{0}$X transition. Figure~\ref{fig:LaserTuningDependence} compares a) the spin noise power and b) the spin dephasing rate of the $\gamma_{h}^{(2)}$ spin noise contribution with the inhomogeneous $\gamma_{i}$ contribution. The spectral position of the maximum spin noise power at 3.360~eV indicates that the major spin noise contribution originate for both from Al-donor bound electrons. However, the total spin noise power of the $\gamma_{h}^{(2)}$ contribution is about one order of magnitude larger than the inhomogeneous contribution. The relative spin noise power scales with the contributing densities of spins such that we conclude, that roughly 90\% of all donor-bound electrons undergo a fast spin-relaxation via defect mediated spin-scattering in conjunction with spin orbit splitting. The exact mechanism is not clear at this point, but we assume an Elliot-Yafet like spin dephasing mechanism where a high scattering rate leads to a higher probability of an electron spin flip \cite{Elliott.PR.1954,Yafet.J.Phys.Chem.Solids.1956}. Under the assumption that at least one defect within the donor volume already leads to a fast spin decay we estimate a defect density $n_{d}$ by relating the donor volume $V_{\rm{Al}}$ to the average defect volume $V_d=n_{d}^{-1} = V_{\rm{Al}}/0.9$ which results in $n_{d}= 3\times 10^{19}\rm{cm}^{-3}$. The high defect density is consistent with the nanocrystalline structure as measured by scanning electron microscopy. Furthermore, the spin dephasing rate $\gamma_{h}^{(2)}$ is about one order of magnitude larger compared to $\gamma_{i}$. In fact it is even higher than $\nu_{\rm{L}}$ for all measured magnetic fields (excluding the background acquisition). This in turn explains the non-Gaussian form of the spin noise contribution associated with $\gamma_{h}^{(2)}$ since the spin relaxation follows that of an overdamped oscillator, which can be approximated by an exponential, i.e., homogenous decay in the time domain. 

In conclusion, we present the first all optical spin noise measurements on the wide-band gap semiconductor material ZnO. All measured spin noise contributions are identified to originate from the Al-donor bound electron in thermal equilibrium. The presented spin noise measurements unveil the rich physics of spin dynamics governed by the Overhauser nuclear field with the extracted inhomogeneous spin dephasing times being in accordance with the peculiar natural ZnO isotope composition. Most interestingly, we found additionally a strong defect mediated spin noise signal giving rise to very short spin lifetimes possibly originating from an Elliot-Yafet like spin flip mechanism but which certainly deserves more extensive future investigations. 

We thank H. Schmidt for taking the SEM pictures and M. Mansur-Al-Suleiman for the helpful discussion. We gratefully acknowledge financial support by the NTH school \textit{Contacts in Nanosystems}, the BMBF project \textit{QuaHLRep} and by the Deutsche Forschungsgemeinschaft in the framework of the priority program SPP1285 \textit{Semiconductor Spintronics}. 

\bibliographystyle{apsrev4-1}   

\end{document}